\documentclass[twocolumn,showpacs,preprintnumbers,amsmath,amssymb,superscriptaddress]{revtex4}
\usepackage{amssymb}
\usepackage{graphicx,color}
\usepackage{bm}

\begin{document}

\title{Jets, Mach cone, hot spots, ridges, harmonic flow, dihadron and $\gamma$-hadron correlation in high-energy heavy-ion collisions}

\author{Guo-Liang Ma}
\affiliation{Shanghai Institute of Applied Physics, Chinese
Academy of Sciences, P.O. Box 800-204, Shanghai 201800, China}
\author{Xin-Nian Wang}
\affiliation{Institute of Particle Physics, Huazhong Normal University, Wuhan 430079, China}
\affiliation{Nuclear Science Division MS 70R0319, Lawrence Berkeley National Laboratory, Berkeley, California 94720}

\begin{abstract}
Within the AMPT Monte Carlo model, fluctuations in the initial transverse parton density are shown 
to lead to harmonic flows. The net back-to-back dihadron azimuthal correlation after subtraction of contributions from harmonic 
flows still has a double-peak that is independent of the initial geometric triangularity
and unique to the jet-induced Mach cone and expanding hot spots distorted by radial flow. The longitudinal structure of hot spots also leads to a near-side ridge in dihadron correlation with a 
large rapidity gap. By successively randomizing the azimuthal angle of the transverse momenta and positions of 
initial partons, one can isolate the effects of jet-induced medium excitation and expanding hot spots on the
dihadron azimuthal correlation. 
The double-peaks in the net dihadron and $\gamma$-hadron
correlation are quantitatively different since the later is caused only by jet-induced Mach cone.
\end{abstract}


\maketitle

Dihadron correlation in high-energy heavy-ion experiments provides a useful tool to study jet quenching \cite{Wang:1991xy}
that suppresses not only single inclusive large $p_{T}$ hadrons \cite{Adler:2003qi} 
but also large $p_{T}$ back-to-back correlation \cite{Adler:2002tq}.
By decreasing $p_{T}$ of the away-side hadrons associated with a high $p_{T}$ triggered hadron, 
one can further study the medium modification of hadron distribution from an
energetic jet as well as the jet-induced medium excitations. Experimental data in $Au+Au$
collisions at the Relativistic Heavy-Ion Collider (RHIC) \cite{Adams:2005ph,Adler:2005ee,:2008nda}
show a  double-peaked back-to-back azimuthal dihadron correlation with a maximum opening 
angle of $\Delta\phi\approx 1$ (rad) relative to the back-side of a high $p_{T}$ trigger.
Such a feature in dihadron correlation was suspected to be caused by Mach-cone
excitation induced by jet-medium interaction \cite{CasalderreySolana:2004qm,Stoecker:2004qu}.
However, hydrodynamic study of jet-induced medium excitation with a realistic source term for
energy-momentum deposition \cite{Betz:2009su} and parton transport study of
jet propagation \cite{Li:2010ts} both show that it is the deflection of jet showers and medium excitation by
radial flow that produces double-peaked back-to-back dihadron correlation.  Moreover, recent
studies also found that expanding hot spots in the initial 
parton transverse density distorted by radial flow \cite{Takahashi:2009na,Shuryak:2009cy} 
and the triangular flow \cite{Alver:2010gr,Schenke:2010rr} all contribute to a double-peaked
back-to-back dihadron correlation. If these local fluctuations are extended in the longitudinal
direction like a string model \cite{Werner:2010aa} or glasma \cite{Gelis:2008ad,Gavin:2008ev}, they would also 
lead to a ridge structure in the near-side dihadron
correlation with large rapidity gap \cite{Adams:2004pa,:2009qa}. With these different mechanisms
contributing to the dihadron correlation, it is important to explore ways to separate different contributions
and study the characteristics of the dihadron correlation from each of them.

In this Letter, we will use a multiphase transport (AMPT) model \cite{Zhang:1999bd} to study dihadron correlation
as a result of harmonic flows, 
expanding hot spots, jets and jet-induced medium excitation.
We first study harmonic flows of hadron spectra due to fluctuations in initial parton density, which all contribute
to dihadron azimuthal correlation. The dihadron correlation after subtraction of contributions
from harmonic flows should provide information on the unique structures of hadron correlation
from medium modified jets,  jet-induced medium excitation and expanding hot spots distorted by
radial flow in high-energy heavy-ion collisions. We will also investigate the longitudinal structures
of hot spots and the resulting dihadron azimuthal correlation with large rapidity gap.  By successively
randomizing the azimuthal angles of transverse momenta and coordinates of initial jet shower
partons, we can isolate the effects of medium modified dijets, jet-induced medium excitation and
expanding hot spots. We will also compare dihadron correlation after subtraction of the contributions
from harmonic flow with $\gamma$-hadron correlation which is only affected
by jet-induced medium excitation. The AMPT results shown in this paper are central $Au+Au$
collisions with fixed impact-parameter $b=0$ at the RHIC energy $\sqrt{s}=200$ GeV.

AMPT \cite{Zhang:1999bd} model combines initial conditions from HIJING model \cite{Wang:1991hta}
with parton and hadron cascade for final state interaction. We will use a version of AMPT with string melting
and  parton recombination for hadronization which was shown to better describe the collective phenomenon
in heavy-ion collisions at RHIC \cite{Lin:2004en}. The parton cascade employed in AMPT includes only elastic 
parton collisions whose cross sections are controlled by the values of strong coupling constant
and the Debye screening mass.  Within HIJING, Glauber model for multiple nucleon scattering 
is used to describe the initial parton production in heavy-ion collisions. Nucleon-nucleon
scatterings contain both independent hard
parton scattering and coherent soft interaction that is modeled by string formation for each
participant nucleon. Strings are then converted into soft partons via string melting scheme
in AMPT. Such multiple scatterings lead to fluctuation in local parton number density or hot spots 
from both soft and hard interactions which are proportional to local transverse density of participant 
nucleons and binary collisions, respectively. 

\begin{figure}
\centerline{\includegraphics[width=10.5cm]{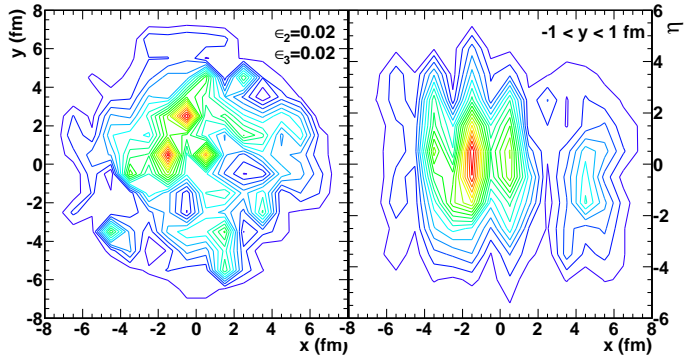}}
\caption{(Color online) Contour plot of initial parton density (in arbitrary unit)  $dN/dxdy$ in transverse
plane (left panel)  and  $dN/dxd\eta$ (right panel) in $x$-$\eta$ (pseudorapidity) plane in a typical AMPT 
central $Au+Au$ event ($b=0$) at $\sqrt{s}=200$ GeV, with ellipticity $\epsilon_{2}=0.02$ 
and triangularity $\epsilon_{3}=0.02$ of the transverse parton distribution. }
 \label{fig-contour}
\end{figure}

Shown in Fig.~\ref{fig-contour} are contour plots of initial parton density in the transverse
plane $dN/dxdy$ (left panel) and in $x-\eta$(pseudorapidity) plane $dN/dxd\eta$ 
within a slice $|y|<1$ fm (right panel) of a typical AMPT event.
We assume the transverse position of both hard and soft partons are
confined to the size of their parent nucleons $r_{N}\sim 1$ fm. Therefore, fluctuations in local density of participant
nucleons and binary nucleon-nucleon collisions lead to fluctuations in parton transverse density or hot spots 
with the smallest transverse size of about 1 fm.
These hot spots and valleys give rise to finite values of initial geometric 
irregularities $\epsilon_{n}$ defined as
\begin{equation}
\epsilon _n  = \frac{{\sqrt {\left\langle {r^2 \cos (n\varphi )} \right\rangle ^2  
+ \left\langle {r^2 \sin (n\varphi)} \right\rangle ^2 } }}{{\left\langle {r^2 } \right\rangle }}, 
\label{eq:epsilon}
\end{equation}
where ($r$, $\varphi$) are polar coordinates of each parton and the average $\langle \cdots\rangle$ 
is density weighted. Normally, $\epsilon_{2}$ is referred to as eccentricity  and $\epsilon_{3}$ as triangularity.

Hot spots in the fluctuating initial parton density distribution are also extended in the longitudinal direction
as shown in the right panel of Fig.~\ref{fig-contour}. Such extended longitudinal distribution in
pseudorapidity $\eta$ is partially from soft partons via the materialization of strings. Partons
from initial state radiation associated with hard scatterings have a distribution $dN/dy=dN/d\log(1/x)\sim 1$ which is
also extended in rapidity.

\begin{figure}
\centerline{\includegraphics[width=8.5cm]{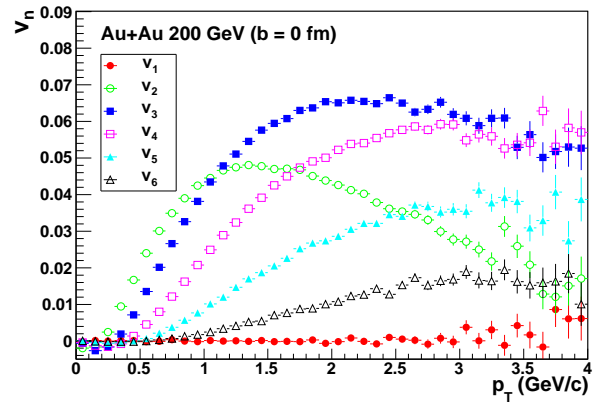}}
\caption{(Color online) Azimuthal anisotropies of hadron spectra $v_{n}(p_{T})$ $(n=1-6)$
in central ($b=0$) $Au+Au$ collisions at $\sqrt{s}=200$ GeV from AMPT model
calculation.}
 \label{fig-vn}
\end{figure}

Collective expansion due to parton rescattering will translate the initial geometric irregularities
into harmonic flows in momentum space \cite{Qin:2010pf}. Shown in Fig.~\ref{fig-vn}
are the harmonic flows of final hadron spectra $v_n  = \left\langle {\cos n (\phi  - \psi _n )} \right\rangle $
from AMPT calculations, where the event plane angle for each harmonics is given by 
\begin{equation}
\psi _n  = \frac{1}{n}\left[ \arctan\frac{\left\langle {r^2 \sin (n\varphi)} \right\rangle}{\left\langle {r^2 \cos (n\varphi)} \right\rangle} 
+ \pi \right].
 \label{eq:psin}
\end{equation}
Even in central collisions at fixed impact-parameter $b=0$, the geometrical fluctuation of initial
parton density leads to
anisotropic collective expansion which translates the initial geometric irregularities
into significant values of harmonic flows in momentum space in the central rapidity region. 
It is interesting to observe that all $v_{n}(p_{T})$ decrease at high $p_{T}$ but the turning points 
shift to higher $p_{T}$ for higher harmonics. The higher harmonic flows $v_{n}$ for $n>6$ are 
insignificant due to viscous diffusion.

\begin{figure}
\centerline{\includegraphics[width=8.5cm]{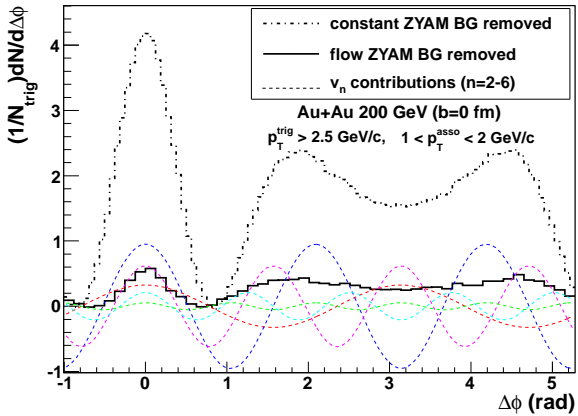}}
\caption{(Color online) AMPT results on dihadron azimuthal correlation before (dot-dashed) and after (solid)
subtraction of contribution from harmonic flow $v_{n} (n=2-6)$ for $p_{T}^{\rm trig}>2.5$ GeV/$c$ 
and $1 < p_{T}^{\rm asso} < 2$ GeV/$c$.}
 \label{fig-dih1}
\end{figure}

In the study of dihadron correlation to search for effects of medium modification of
jet structure and jet-induced medium excitations, it is important to isolate and subtract
contributions from harmonic flows, especially the triangular flow which
contributes the most to the double-peak structure of back-to-back dihadron
correlation. We will use the ZYAM (zero yield at minimum) scheme \cite{Adams:2005ph}
to subtract contributions to dihadron correlation,
\begin{equation}
f(\Delta \phi ) = B\left(1 + \sum\limits_{n = 1}^\infty {2\langle v_n^{\rm trig} v_n^{\rm asso}\rangle \cos n\Delta \phi} \right),  \label{eq:BG}
\end{equation}
 from harmonic flows, where B is a normalization factor determined by the ZYAM scheme, $v_n^{trig}$ and $v_n^{asso}$ 
are harmonic flow coefficients for trigger and associated hadrons.  Shown in Fig.~\ref{fig-dih1}
are dihadron correlations before (dot-dashed) and after (solid) the removal of
contributions from harmonic flows for $p_{T}^{\rm trig}>2.5$ GeV/$c$ 
and $1 < p_{T}^{\rm asso} < 2$ GeV/$c$.
Also shown are contributions from each harmonic flow $n=2$-6 (dashed). 
These contributions are significant for up to $n=5$ harmonics.

\begin{figure}
\centerline{\includegraphics[width=8.5cm]{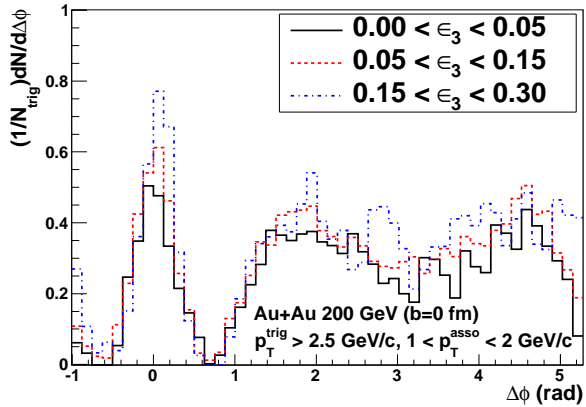}}
\caption{(Color online) Dihadron correlations after subtraction of harmonic flow
with different values of geometric triangularity $\epsilon_{3}$ for $p_{T}^{\rm trig}>2.5$ GeV/$c$ 
and $1 < p_{T}^{\rm asso} < 2$ GeV/$c$ .}
 \label{fig-dih2}
\end{figure}

After subtraction of contributions from harmonic flows, the dihadron correlation still has
a double-peak feature on the away-side of the trigger which should reflect the azimuthal structure 
from deflection of medium modified dijets and jet-induced medium excitations and hadrons 
from expanding hot spots under strong radial flow. The structure therefore should
be unique and insensitive to the fluctuation of the initial geometry of dense matter at a fixed
impact-parameter. As shown in Fig.~\ref{fig-dih2}, the dihadron correlations after subtraction of
contributions from harmonic flows become independent on the initial geometric
triangularity $\epsilon_{3}$.  

\begin{figure}
\centerline{\includegraphics[width=8.5cm]{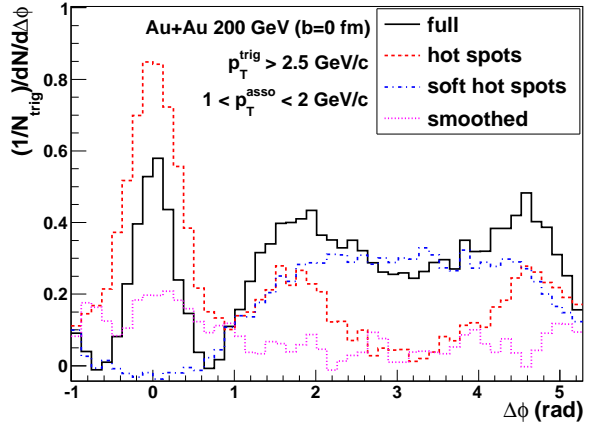}}
\caption{(Color online) Dihadron correlation (with harmonic flow subtracted) from AMPT with different
initial conditions for $p_{T}^{\rm trig}>2.5$ GeV/$c$ and $1 < p_{T}^{\rm asso} < 2$ GeV/$c$. 
See text for details on the different initial conditions.}
 \label{fig-dih3}
\end{figure}

\begin{figure}
\centerline{\includegraphics[width=8.0cm]{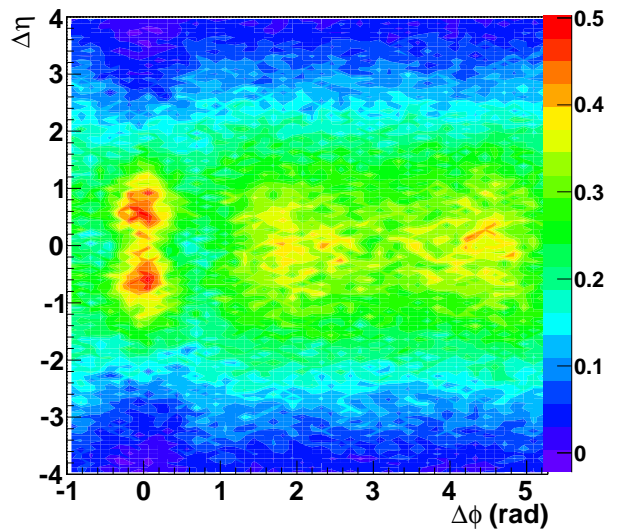}}
\caption{(Color online) The contour plot of dihadron correlation (with harmonic flow subtracted) 
from AMPT in azimuthal and pseudorapidity for $p_{T}^{\rm trig}>2.5$ GeV/$c$ and $1 < p_{T}^{\rm asso} < 2$ GeV/$c$.}
 \label{fig-ridge}
\end{figure}

In order to study the structure of dihadron azimuthal correlation from jets (including jet-induced medium
excitation) and hot spots separately, we successively switch off each mechanism and calculate the 
dihadron correlation within AMPT model. We first randomize the azimuthal angle of each jet shower
parton in the initial condition from HIJING simulations. This effectively switches off the initial
back-to-back correlation of dijets. The dihadron correlation (dashed) denoted as ``hot spots'' in Fig.~\ref{fig-dih3}
still exhibits a double-peak on the away-side that comes only from hot spots. It has roughly the 
same opening angle $\Delta\phi\sim 1$ (rad) as in the ``full'' simulation (solid).  However, the magnitude
of the double-peak in the away-side correlation is reduced by about 40\%, which can be attributed to
dihadrons from medium modified dijets and jet-induced medium excitation. Without knowing
the relative yields of hadrons from jets and hot spots, it is difficult to extract dihadron correlation
from dijets (and jet-induced medium excitation) alone. When jet production is turned off in the HIJING
initial condition,  fluctuation in soft partons from strings can still form  what we denote as ``soft hot spots''
that lead to a back-to-back dihadron correlation (dot-dashed) with a weak
double-peak. It is clear that jet shower partons increase the local parton density in ``hot spots'' and lead
to a stronger double-peaked dihadron correlation than that of ``soft hot spots''.
Such ``soft hot spots'' are the likely candidate mechanism for the observed back-to-back
dihadron correlation in heavy-ion collisions at the SPS energy \cite{Ploskon:2005pa} where jet production
is insignificant. Without jets in AMPT, one can further randomize the polar angle of  transverse coordinates of
soft partons and therefore eliminate the ``soft hot spots''. The dihadron correlation from such ``smoothed'' initial
condition becomes almost flat (dotted).

Because of the elongated shape of hot spots in the longitudinal direction
as shown in Fig.~\ref{fig-contour} (left panel), the structure of dihadron azimuthal correlation should
remain similar with a large range of rapidity gaps as shown in Fig.~\ref{fig-ridge}.  Since dihadron correlation
from a single jet is only restricted to small rapidity gap $\delta\eta\sim 1$, the near-side azimuthal
correlation with a large rapidity gap, referred to as ridge,  can only come from a hot spot.
This is confirmed by our observation that dihadron
correlation in AMPT with the ``smoothed'' initial conditions does not have any ridge structure.
In addition, jet production is approximately proportional to binary nucleon-nucleon collisions and
should also be correlated with hot spots. Therefore, the ridge in dihadron correlation can occur for 
both hard and soft trigger hadrons. In non-central collisions, the effects of jet-medium interaction and
hot spots are shown to give only a broad single peak in dihadron correlation on the away-side \cite{Xu:2010du}.
With a large rapidity gap, it amounts to  a dihadron correlation due to momentum conservation \cite{Luzum:2010sp}.

\begin{figure}
\centerline{\includegraphics[width=8.5cm]{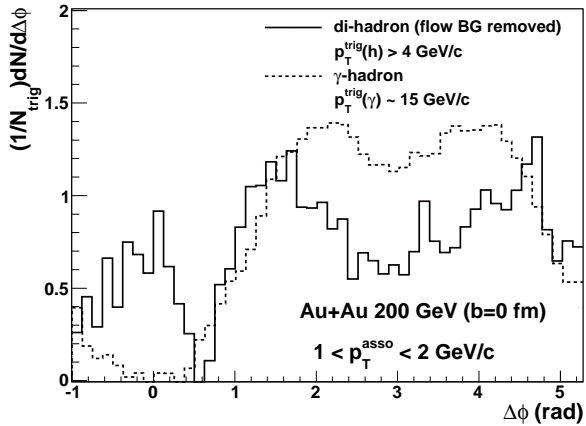}}
\caption{Dihadron correlation (solid)
compared with $\gamma$-hadron correlation (dashed) from AMPT for $p_{T}^{\rm trig}(h)>4$ GeV/$c$,
$p_{T}^{\rm trig}(\gamma)\ge 15$ GeV/$c$ and $1 < p_{T}^{\rm asso} < 2$ GeV/$c$ .}
 \label{fig-gamma}
\end{figure}

Since hard direct photons are produced uniformly in azimuthal angle, any structure in $\gamma$-hadron 
azimuthal correlation with large $p_{\rm trig}^{\gamma}$ can only come from $\gamma$-triggered jets
and jet-induced medium excitation \cite{Li:2010ts}. Shown in Fig.~\ref{fig-gamma} are dihadron correlations
(solid) after subtraction of harmonic flows as compared with $\gamma$-hadron
correlation (dashed). The two correlations are comparable in magnitude but dihadron has a more pronounced 
double-peak which can be attributed to additional dihadrons from hot spots and the
geometric bias toward surface and tangential emission that enhances deflection
of jet showers and jet-induced medium excitation \cite{Li:2010ts} by radial flow.
Such difference is important to measure in experiments
that will provide critical information on jet-induced medium excitation and hydrodynamic evolution of hot
spots in high-energy heavy-ion collisions.


This work is supported by the NSFC of China under Projects Nos. 10705044, 10975059, 11035009,
the Knowledge Innovation Project of Chinese Academy of Sciences under Grant No. KJCX2-EW-N01
and by the U.S. DOE under Contract No. DE-AC02-05CH11231 and within the framework of the JET Collaboration.

\end{document}